\documentstyle[aps,preprint]{revtex}
\begin{document}

\draft
\title{Chern-Simons vortices in an open system}
\author{Mark Burgess}
\address{Faculty of Engineering, Oslo College, 0254 Oslo,
Norway\\and\\Institute of Physics, University of Oslo, P.O.Box 1048 Blindern,
0316 Oslo, Norway}
\date{\today}

\bibliographystyle{unsrt}
\maketitle

\begin{abstract}
A gauge invariant quantum field theory with a spacetime dependent
Chern-Simons coefficient is studied. Using a constraint formalism
together with the Schwinger action principle it is shown that
non-zero gradients in the coefficient induce
magnetic-moment corrections to the
Hall current and transform vortex singularities into non-local
objects. The fundamental commutator for the density fluctuations
is obtained from the action principle and the Hamiltonian of the
Chern-Simons field is shown to vanish only under the restricted class
of variations which satisfy the gauge invariance constraint.
\end{abstract}

The study of Chern-Simons theories is motivated principally by two
observations, namely that important aspects of the quantum Hall
phenomenon are described efficiently by a Chern-Simons theory, and
that a viable theory of high temperature superconductivity should be
characterized by a parity-violating antiferromagnetic state\cite{back}.
Symmetry
considerations alone suggest that such an interaction
should be present in these systems.

In the case of the superconductor, where electrons are effectivly
two dimensional by virtue of the layered symmetry, neighbouring planes
can be expected to play a non-trivial role on the dynamics of the two
dimensional system. In particular, donor sites and irregularities
in neighbouring two-dimensional systems could have a sufficiently
coherent influence on a two dimensional system that the physical
properties in the two dimensional superconductor are modulated by
the presence of their neighbours. This would suggest an
{\em effective} field theory with position dependent couplings.
In a similar vein, it was suggested by Jacobs\cite{jacobs1} that certain
desirable features
might be achieved if the Chern-Simons term was coupled, not by a
coupling constant, but through an `axion' field -- i.e. a spacetime
dependent coupling. In a continuum theory of the quantum Hall effect,
a stepping Chern-Simons coefficient is also natural in the vicinity of
the edges of the Hall sample where the statistics parameter passes
through a sequence of values dictated by the Landau level structure.
Recent work by the author\cite{burgess1,burgess2} has lead to a formalism
for dealing with the apparent inconsistencies in the interpretation of
such a theory. Although originally motivated on other grounds, the formalism is
easily
adapted to the problem of Chern-Simons particles (the anyon system\cite{back})
which has been investigated in refs. \cite{jacobs1,jackiw1}.

The apparent difficulty with a variable Chern-Simons coefficient is that
the resulting theory is not explicitly gauge invariant. One might argue
that this is because one starts with the action $S$ which is not
a physical object. One could, after all, simply start with the field
equations and make the Chern-Simons coefficient spacetime dependent.
However, in present day quantum field theory the action is increasingly
regarded as being a physical object -- not only its variation:
the Chern-Simons term is a case in point. It is therefore important
to secure a formalism which guarantees consistency between variations
of the action and the dynamical structure of the theory at all levels.
Such a formalism was recently constructed and the physical
meaning of the procedure identified as being that of closing an
open physical system through the use of a constraint. The formalism
is easily adapted to the quantized theory by adopting Schwinger's
action principle. Let us therefore begin by examining the formalism.

The fundamental relation in Schwinger's quantum action principle is
\begin{equation}
\delta \langle t'|t\rangle=i\langle t'|\int_t^{t'}L(q)\, dt|t\rangle.
\end{equation}
{}From this relation
one infers both the operator equations of motion $\frac{\delta S}{\delta q}=0$,
for dynamical variables $q$ and the generator of infinitesimal unitary
transformations $G$ which is obtained from the total time dervitative
in $\delta S$. $S$ is an action symmetrized with respect
to the kinematical derivatives
of the dynamical variables. From this, one obtains the variation of any
operator $A$ on the basis $|t\rangle$
\begin{equation}
\delta A = -i [A,G].
\end{equation}
Consider first the usual Chern-Simons theory for constant
$\mu$. This will serve as a point of reference for the remainder
of the paper. It can be noted that the present formalism bears a certain
resemblance to
the Schr\" odinger quantization examined by Dunne et al\cite{dunne1}
and reproduces the relevant results.
The pure Chern-Simons theory is described by the action
\begin{equation}
S = \int dt d^2x \lbrace \frac{1}{2}\mu\epsilon^{\mu\nu\lambda}
A_\mu\partial_\nu A_\lambda-J^\mu A_\mu\rbrace
\end{equation}
where $J^\mu$ is a gauge invariant current operator and
$\mu$ is constant. The variation
of this action operator with respect to $A_\mu$ leads to the operator
equation of motion
\begin{equation}
\frac{1}{2}\mu \epsilon^{\mu\nu\lambda} F_{\nu\lambda}=J^\mu
\end{equation}
and the generator of infinitesimal unitary transformations on the
field variables\cite{foot1}
\begin{equation}
G(\sigma) = \int d\sigma_\mu\;\mu\epsilon^{\mu\nu\lambda}A_\mu \delta
A_\lambda.
\end{equation}
Taking $\sigma$ to be a spacelike hypersurface, with unit normal
parallel to the
time $t$, one obtains the fundamental commutator for $A_\mu$
trivially by considering $\delta A_\mu$ in (2):
\begin{equation}
[A_i(x),A_j(x')]\Big|_{t=t'} = i\mu^{-1} \epsilon_{ij}\delta({\bf x},{\bf x'}).
\end{equation}
No restrictions are placed on the $A_0$ component which is therefore not
a true canonical variable, rather it should be understood
as a Lagrange multiplier
which enforces the relation $\mu B=-\rho$.

The generator $G(\sigma)$ is not obviously gauge invariant but, if one ignores
 the
source $J_\mu$ for a moment, it is clear that the constraint $B=0$ can
be satisfied by $A_i = \partial_i\xi$, for some scalar field $\xi$.
If one uses this in the generator, it is evident that there is no
dynamical evolution unless $[\partial_1,\partial_2]\xi\not=0$. This
indicates that vortex singularities play a special role in this theory
and that a non-trivial generator with $B=0$ could only be satisfied by a
pointlike source $J^\mu$, as in the flux line singularities of anyon theory.

More generally, if one
solves the field equations giving
\begin{equation}
A_i(x)=\frac{\epsilon_{ji}\partial_j\rho(x)}{\mu\nabla^2}
\end{equation}
and uses this to express the gauge field purely in terms of gauge
invariant operators, one obtains an implicit equation for the
commutator of the density operator, thus identifying density fluctuations
as the basic excitations.
\begin{equation}
4\pi\mu i\delta({\bf x},{\bf x'})=\int d^2x'' d^2x''' [\rho(x''),\rho(x''')]
\epsilon_{lj}\frac{({\bf x}-{\bf x''})^l}{|{\bf x}-{\bf x''}|^2}
\frac{({\bf x'}-{\bf x'''})^j}{|{\bf x'}-{\bf x'''}|^2}\Big|_{t'=t}
\end{equation}
or
\begin{equation}
[\rho({\bf x''},t),\rho({\bf x'''},t)]=4\pi^2\mu \Omega^{-1}i
\delta({\bf x''},{\bf x'''})\epsilon_{lj}
\frac{({\bf x}-{\bf x''})^l}{|{\bf x}-{\bf x''}|^2}
\frac{({\bf x}-{\bf x'''})^j}{|{\bf x}-{\bf x'''}|^2}
\end{equation}
where $\Omega$ has the dimensions of volume.
Since the Chern-Simons action is linear in the time derivative, it
possesses no dynamics independently of $J^\mu$ and thus its sole effect is
to induce certain symmetry relations on the field operators, a fact which is
manifest in the above expression. In deriving (9), a number of relations
concerning vortex fluxline singularities have been used. It is convenient to
state these for the record
\begin{eqnarray}
\tan\theta({\bf x}-{\bf x'}) &=& \frac{({\bf x}-{\bf x'})^2}{({\bf x}-{\bf
x'})^1}\\
-\frac{1}{2\pi}(\partial_l\theta)&=&\epsilon_{lj}\partial_j g({\bf x}-{\bf
x'})\\
\nabla^2  g({\bf x}-{\bf x'}) &=&  \delta({\bf x}-{\bf x'})\\
 g({\bf x}-{\bf x'}) &=& \frac{1}{2\pi}\ln |{\bf x}-{\bf x'}|
\end{eqnarray}
$\theta$ is formally the winding angle between two flux singularities and
satisfies the curious relation
\begin{equation}
[\partial_1,\partial_2]\theta({\bf x}-{\bf x'})=2\pi \delta({\bf x}-{\bf x'}).
\end{equation}
These relations will be a useful reference later when interpreting the
equations of motion for the field operators. (Note also the discussion in
\cite{hagen}
concerning these relations.)

Let us now turn to the case in which the coefficient $\mu(x)$ is an
arbitrary function. As shown in ref. \cite{burgess1}, this
necessitates an additional variable coupling to the source in order to
satisfy a suitable gauge invariance constraint:
\begin{equation}
S=\int dV_x \lbrace \frac{1}{2}\mu(x)\epsilon^{\mu\nu\lambda}A_\mu\partial_\nu
A_\lambda
-f(x)J^\mu A_\mu \rbrace.
\end{equation}
Since both couplings are position dependent, this represents a phenomenological
system rather than a fundamental one. In order to proceed, one needs to
apply a physical boundary condition to the source. As explained
earlier\cite{burgess1},
the consistency of this theory
then requires that the source be adjusted in such as a way that
gauge invariance is maintained and energy is conserved. Since we do not
want the source coupling to vanish when $\mu$ is constant, the natural boundary
condition in this instance is $f(x)=\mu(x)/\alpha$, for some
constant mass scale $\alpha$.
Thus, after a convenient rescaling, one may write
\begin{equation}
S=\int dV_x \mu(x) \lbrace \frac{1}{2}\alpha \epsilon^{\mu\nu\lambda}
A_\mu\partial_\nu A_\lambda -J^\mu A_\mu\rbrace
\end{equation}
where $\mu(x)$ is now a dimensionless field. The role of $\mu(x)$ is
to present the system through a `distorting glass'. The physical
picture is that of a two dimensional gas of particles influenced
microscopically but smoothly by sites in neighbouring planar systems.
The special form
of the action together with the constraint results
in the preservation of gauge invariance.

The allowed class of variations of the action is determined
from the consideration of an
infinitesimal gauge transformation $A_\mu\rightarrow A_\mu+\partial_\mu\xi$,
which provides us with an operator constraint.
We shall assume that the current $J^\mu$ is conserved and that the
variation of $\xi$ commutes with the field. On varying the action with
respect to $\delta\xi$, one obtains the constraint
\begin{equation}
\frac{1}{2}\alpha\epsilon^{\mu\nu\lambda}\partial_\nu A_\lambda=J^\mu
\end{equation}
and the generator of infinitesmial gauge transformations
\begin{equation}
G_\xi=\int d\sigma [\frac{1}{2}\alpha\mu\epsilon^{ij}
\partial_j A_i-\mu J^0]\delta\xi.
\end{equation}
These are gauge invariant, indeed one sees how the formalism which
includes the physical boundary condition repairs the
canonical structure of the theory in the presence of variable $\mu(x)$.
The solutions to (17) determine now the class of variations under which the
quantum theory will be gauge invariant. Choosing the Coulomb gauge to
eliminate the unphysical degrees of freedom from the field operators, one
may solve (17) to get
\begin{equation}
A_\sigma = 2\alpha^{-1}\int d^2{\bf x}'
\epsilon_{\sigma\rho\lambda}\partial^\rho
J^\lambda({\bf x'})g({\bf x},{\bf x'}).
\end{equation}
The variation of this result now yields the allowed values
for $\delta A_\mu$. Returning to (16) one may thus vary with respect to the
dynamical variable $A_\mu$ to obtain the gauge invariant equations of
motion for the field operators.
\begin{eqnarray}
J^i&=&\alpha \epsilon_{ij}E^j+\alpha \epsilon^{ij}(\partial_j\mu)\mu^{-1}
\frac{\epsilon_{lm}\partial^lJ^m}{\nabla^2} -
\alpha\epsilon^{ij}(\partial_0\mu)\mu^{-1}\frac{\epsilon_{lj}\partial^l\rho}{\nabla^2}\\
\rho &=& -\alpha
B-\alpha\epsilon^{ij}(\partial_j\mu)\mu^{-1}\frac{\epsilon_{li}\partial^l\rho}{\nabla^2}.
\end{eqnarray}
The first of these equations clearly describes a modification to the
Hall current of the system. The spatial gradient of $\mu$ makes
the current dependent on its own curl in precisely the manner of a magnetic
moment interaction\cite{stern1,burgess2,gaboretc}. It is interesting to compare
this form to the parallel
theory\cite{gaboretc} in which the gauge field couples directly
to the source through a parity violating term. The same magnetic current
loop interaction appears in both cases.
The time gradient term
leads to an additional induction effect.

To ascertain the meaning of the second equation, it is useful to
define a field $\theta$ by analogy with
equation (11). Now, integrating by parts and assuming only weakly varying
$\mu$, one obtains
\begin{equation}
[\partial_1,\partial_2]\theta({\bf x})\sim 2\pi B \sim 2\pi\delta({\bf x})
\end{equation}
since $\rho\sim-\alpha B$. The translational invariance of the
field $\theta$ has also been assumed. This `rough and ready'
last step serves mainly as a guide to physical intuition and
shows that (21) predicts a non-local
generalization of the vortex lines in the theory with constant $\mu$.

Extracting the generator of infinitesimal unitary transformations from
the variation of the action operator, one easily determines that the
commutator analogous to (8) is given by the implicit equation
\begin{equation}
i\alpha\mu^{-1}\pi^2\delta({\bf x},{\bf x'}) = \int d^2{\bf x''} d^2{\bf x'''}
[\rho({\bf x''}),\rho({\bf x'''})]\epsilon_{lj}
\frac{({\bf x}-{\bf x''})^l}{|{\bf x}-{\bf x''}|^2}
\frac{({\bf x'}-{\bf x'''})^j}{|{\bf x'}-{\bf x'''}|^2}.
\end{equation}

Finally, since the Chern-Simons term imparts no dynamics to the system, the
Hamiltonian
must be expected to vanish.
The Hamiltonian for the Chern-Simons action
can be computed from $H = -\frac{\delta S}{\delta t}$
and is indeed found to vanish under the restricted class of variations in (19).
Under general variations, it is non-vanishing when $\mu(x)$ is spacetime
dependent. The time variation may be defined by
\begin{equation}
\delta S = \int_t^{t+\delta t} L(A_\mu,J_\nu) = \int dt\; \delta L,
\end{equation}
where, to first order
\begin{eqnarray}
\delta \mu(x) &=& \frac{\partial\mu}{\partial t}\delta t\\
\delta A_\mu &=& F_\mu^{~\sigma}\delta x_\sigma.
\end{eqnarray}
The latter gauge invariant transformation is required to generate the
symmetrical, conserved energy-momentum tensor for the theory\cite{jackiw2}.
The Hamiltonian operator is therefore
\begin{equation}
H = - \int d^2{\bf x}\;(\partial_t\mu)\epsilon^{\mu\nu\lambda}A_\mu\partial_\nu
A_\lambda.
\end{equation}
On using the solution of the operator equations of motion (19) this is
seen to vanish as required.
The reason has already been described in earlier work: the interpretation
of the naive unconstrained theory is that of
an open system and the energy is therefore not automatically conserved.
One would therefore encounter a non-vanishing Hamiltonian.

An interesting feature of the present vortex system is that the
gauge invariance constraint (17) does not involve the spacetime dependent
field $\mu(x)$ unlike the Maxwell-Chern-Simons theory in ref.
\cite{burgess1,burgess2}. This has an important implication -- namely
that, in the absence of external magnetic fields, the flux lines
can form arbitrary stable gradients in $\mu$ without violating gauge
invariance.
This must be understood as a topological phenomenon since the
relations provide no dynamical reason for such behaviour. It might
be possible in certain cases to identify these with spin textures.
The obvious information we are missing
which decides these gradients is the details of the neighbouring
system(s). One would
expect, on the basis of experience with the Maxwell-Chern-Simons system,
that when the coupling to the external system is removed, the Chern-Simons
coefficient would have to decay to a constant value. This is indeed
the case. If one relaxes the imposed boundary condition and takes
$f(x)\rightarrow const$, then the gauge invariance condition leads to
the familiar equation\cite{burgess1}
\begin{equation}
(\partial_0 \mu) B + (\partial_i\mu)\epsilon^{ij}E_j=0
\end{equation}
which has decaying solutions in the manner of the Langevin equation.
Thus the interpretation of the system is fully self-consistent.

To summarize, a Chern-Simons field theory coupled to a gauge invariant
current $J^\mu$ through the field $\mu(x)$ is only gauge invariant
and unitary under a restricted class of operator variations. This
can be understood as arising from an interaction with an external system.
The restricted theory can be explored with the help of a constraint formalism
applied to the Schwinger action principle. The corrections to
regular Chern-Simons theory indicate a modification of the Hall
current for vortex lines in a manner which resembles a magnetic
moment interaction term and an induction term. The sharp nature
of the vortices is distorted by the gradients in $\mu(x)$ but the
basic excitations are of a similar nature.

It should be possible, by supplementing the source terms with extra
impulsive sources, to compute the many body Green functions for this
theory directly from the Schwinger action principle. These may then
be used to determine the corrections to the thermodynamical and
transport properties of this model, particularly
the effect of the gradients in $\mu(x)$ on the conductivity
in a model for a superconductor. The present results are model independent,
but agree well with the specific
model presented in ref. \cite{jacobs1} and back up the work of
ref. \cite{burgess2}.

The present model, motivated essentially by symmetry considerations
and its connection with the widely discussed anyon model,
has been simplified as far as possible for the sake of illustration.
A more realistic model would be more specific about the origin
of the source terms and must provide some empirical estimate of
the strength of the coupling, perhaps using data for the observed
magnetic moment interactions in high $T_c$ superconductors.
These points turn out to involve some subtle issues and will be pursued
elsewhere.


\begin{thebibliography}{99}
\bibitem{foot1} Note that the definition of the generator differs from
Schwinger's definition
by a factor of two in the case of the linear time-derivative. This
is done so that the
unitary evolution rule (2) is general as it should be.
\end{thebibliography}
\end{document}